\documentclass[twocolumn]{aastex631}

\newcommand{\kms}{\mbox{km s$^{-1}$}}
\newcommand{\x}{\mbox{$\times$}}
\newcommand{\Msun}{\mbox{$M_{\odot}$}}
\newcommand{\Lsun}{\mbox{$L_{\odot}$}}

\shorttitle{Gravity Driven Magnetic Field at $\sim$1000 au Scales in High-mass Star Formation}
\shortauthors{Sanhueza et al.}
\graphicspath{{./}{figures/}}

\begin{document}

\title{Gravity Driven Magnetic Field at $\sim$1000 au Scales in High-mass Star Formation}

\correspondingauthor{Patricio Sanhueza}
\email{patricio.sanhueza@nao.ac.jp}

\author[0000-0002-7125-7685]{Patricio Sanhueza}
\affiliation{National Astronomical Observatory of Japan, National Institutes of Natural Sciences, 2-21-1 Osawa, Mitaka, Tokyo 181-8588, Japan}
\affiliation{Department of Astronomical Science, SOKENDAI (The Graduate University for Advanced Studies), 2-21-1 Osawa, Mitaka, Tokyo 181-8588, Japan}

\author{Josep Miquel Girart}
\affiliation{Institut de Ciencies de l'Espai (ICE, CSIC), Can Magrans s/n, 08193, Cerdanyola del Vallès, Catalonia, Spain}
\affiliation{Institut d'Estudis Espacials de Catalunya (IEEC), 08034, Barcelona, Catalonia, Spain}

\author{Marco Padovani}
\affiliation{INAF–Osservatorio Astrofisico di Arcetri, Largo E. Fermi 5, 50125 Firenze, Italy}

\author{Daniele Galli}
\affiliation{INAF–Osservatorio Astrofisico di Arcetri, Largo E. Fermi 5, 50125 Firenze, Italy}

\author{Charles L. H. Hull}
\affiliation{National Astronomical Observatory of Japan, Alonso de C\'ordova 3788, Office 61B, 7630422, Vitacura, Santiago, Chile}
\affiliation{Joint ALMA Observatory, Alonso de C\'ordova 3107, Vitacura, Santiago, Chile}
\affiliation{NAOJ Fellow}

\author{Qizhou Zhang}
\affiliation{Center for Astrophysics $|$ Harvard \& Smithsonian, 60 Garden Street, Cambridge, MA 02138, USA}

\author{Paulo Cortes}
\affiliation{National Radio Astronomy Observatory, 520 Edgemont Road, Charlottesville, VA 22903, USA}

\author{Ian W. Stephens}
\affiliation{Department of Earth, Environment and Physics, Worcester State University, Worcester, MA 01602, USA}

\author{Manuel Fern\'andez-L\'opez}
\affiliation{Instituto Argentino de Radioastronom\'ia (CCT-La Plata, CONICET; CICPBA), C.C. No. 5, 1894, Villa Elisa, Buenos Aires, Argentina}

\author{James M. Jackson}
\affiliation{USRA SOFIA Science Center, NASA Ames Research Center, Moffett Field, CA 94045, USA}

\author{Pau Frau}
\affiliation{Private researcher}

\author{Patrick M. Kock}
\affiliation{Academia Sinica Institute of Astronomy and Astrophysics, 11F of Astro-Math Bldg, 1, Section 4, Roosevelt Road, Taipei 10617, Taiwan}

\author{Benjamin Wu}
\affiliation{NVIDIA Research, 2788 San Tomas Expressway, Santa Clara, CA 95051, USA}
\affiliation{National Astronomical Observatory of Japan, National Institutes of Natural Sciences, 2-21-1 Osawa, Mitaka, Tokyo 181-8588, Japan}

\author{Luis A. Zapata}
\affiliation{Instituto de Radioastronom\'ia y Astrof\'isica, Universidad Nacional Aut\'onoma de M\'exico, P.O. Box 3-72, 58090, Morelia, Michoac\'an, Mexico}

\author{Fernando Olguin}
\affiliation{Institute of Astronomy and Department of Physics, National Tsing Hua University, Hsinchu 30013, Taiwan}

\author{Xing Lu}
\affiliation{National Astronomical Observatory of Japan, National Institutes of Natural Sciences, 2-21-1 Osawa, Mitaka, Tokyo 181-8588, Japan}

\author{Andrea Silva}
\affiliation{Kavli Institute for the Physics and Mathematics of the Universe, The University of Tokyo, Kashiwa, 277-8583 (Kavli IPMU, WPI) Japan}
\affiliation{Department of Astronomy, School of Science, The University of Tokyo, 7-3-1 Hongo, Bunkyo, Tokyo 113-0033, Japan}
\affiliation{National Astronomical Observatory of Japan, National Institutes of Natural Sciences, 2-21-1 Osawa, Mitaka, Tokyo 181-8588, Japan}

\author{Ya-Wen Tang}
\affiliation{Academia Sinica Institute of Astronomy and Astrophysics, 11F of Astro-Math Bldg, 1, Section 4, Roosevelt Road, Taipei 10617, Taiwan}

\author{Takeshi Sakai}
\affiliation{Graduate School of Informatics and Engineering, The University of Electro-Communications, Chofu, Tokyo 182-8585, Japan}

\author{Andr\'es E. Guzm\'an}
\affiliation{National Astronomical Observatory of Japan, National Institutes of Natural Sciences, 2-21-1 Osawa, Mitaka, Tokyo 181-8588, Japan}

\author{Ken'ichi Tatematsu}
\affiliation{National Astronomical Observatory of Japan, National Institutes of Natural Sciences, 2-21-1 Osawa, Mitaka, Tokyo 181-8588, Japan}
\affiliation{Department of Astronomical Science, SOKENDAI (The Graduate University for Advanced Studies), 2-21-1 Osawa, Mitaka, Tokyo 181-8588, Japan}

\author{Fumitaka Nakamura}
\affiliation{National Astronomical Observatory of Japan, National Institutes of Natural Sciences, 2-21-1 Osawa, Mitaka, Tokyo 181-8588, Japan}
\affiliation{Department of Astronomical Science, SOKENDAI (The Graduate University for Advanced Studies), 2-21-1 Osawa, Mitaka, Tokyo 181-8588, Japan}

\author{Huei-Ru Vivien Chen}
\affiliation{Institute of Astronomy and Department of Physics, National Tsing Hua University, Hsinchu 30013, Taiwan}




\begin{abstract}

A full understanding of high-mass star formation requires the study of one of the most elusive components of the energy balance in the interstellar medium: magnetic fields. We report ALMA 1.2 mm, high-resolution (700 au) dust polarization and molecular line observations of the rotating hot molecular core embedded in the high-mass star-forming region IRAS 18089$-$1732. The dust continuum emission and magnetic field morphology present spiral-like features resembling a whirlpool.  The velocity field traced by the H$^{13}$CO$^+$ (J=3-2) transition line reveals a complex structure with spiral filaments that are likely infalling and rotating, dragging the field with them. We have modeled the magnetic field and find that the best model corresponds to a weakly magnetized core with a  mass-to-magnetic-flux  ratio ($\lambda$) of 8.38. 
The modeled magnetic field is dominated by a poloidal component, but with an important contribution from the toroidal component that has a magnitude of 30\% of the poloidal component. Using the Davis-Chandrasekhar-Fermi method, we estimate a magnetic field strength of 3.5 mG. At the spatial scales accessible to ALMA, an analysis of the energy balance of the system indicates that gravity overwhelms turbulence, rotation, and the magnetic field. We show that high-mass star formation can occur in weakly magnetized environments, with gravity taking the dominant role. 

\end{abstract}



\section{Introduction} \label{sec:intro}

High-mass stars dominate the energy input and chemical enrichment of galaxies. Among all the ingredients that have to be considered in their formation, the magnetic field is by far the least explored. Indeed, it is still debated how the magnetic energy compares to other energies in play, namely turbulence, gravity, and rotation. 

Observations of  linearly polarized dust emission are currently the best available tool to infer the magnetic field in molecular clouds and denser regions  associated with star formation \citep[e.g.,][]{Hull19}. Dust polarization observations of high-mass star-forming regions suggest that the magnetic field appears to be dynamically important during the collapse and fragmentation of parsec-scale molecular clumps and the mass assembly of dense cores at scales of 0.01 to 0.1 pc \citep{Zhang14,Hull19,Cortes19}. However, it remains unclear whether at smaller scales (core-disk interface, $\sim$1000 au) the star formation process is magnetically dominated 
\citep{Girart09,Beltran19,Beuther20,Cortes21,Fernandez21}. So far, since polarization observations are scarce at these small scales, it is  difficult to evaluate the overall importance of the magnetic field in the high-mass star formation process. 

Located at a parallax distance of 2.34 kpc \citep{Xu11} with a bolometric luminosity of 1.3 $\times$ 10$^4$ \Lsun\ \citep{Sridharan02}, the high-mass star-forming region IRAS 18089$-$1732 is an ideal laboratory to assess the importance of the magnetic field with respect to turbulence, gravity, and rotation. Earlier studies at arcsec resolution show that IRAS 18089$-$1732 has a deeply embedded hot core \citep{Beuther04a,Beuther04b} and a disk-like rotating structure roughly perpendicular to a molecular outflow. A line-of-sight magnetic field strength of 8.4 and 5.5 mG has been estimated from measurements of  Zeeman splitting of the 6.7 GHz CH$_3$OH maser line using the 100 m Effelsberg telescope \citep{Vlemmings08} and the MultiElement Radio Linked Interferometer Network, MERLIN  \citep{Olio17}, respectively. Early observations of IRAS 18089$-$1732 with the Submilimeter array (SMA) show the detection of at least a few independents polarization measurements in dust continuum emission \citep{Beuther10}. Taking advantage of the superlative capabilities of the Atacama Large Millimeter/submillimeter Array (ALMA), we have observed IRAS 18089$-$1732 in polarized dust continuum and molecular line emission to better understand the role of the magnetic field in the formation of high-mass stars. This target was observed as part of the Magnetic fields in Massive star-forming Regions (MagMaR) survey that in total contains 30 sources. Details on the survey and source selection will be given in Sanhueza et al. (2021, in prep.). Early results on two high-mass star-forming regions,  G5.89–0.39 and NGC 6334I(N), are  presented in \cite{Fernandez21} and Cortes et al. (2021, submitted), respectively. 

\section{Observations} \label{sec:obs}

\begin{figure*}[ht!]
\plotone{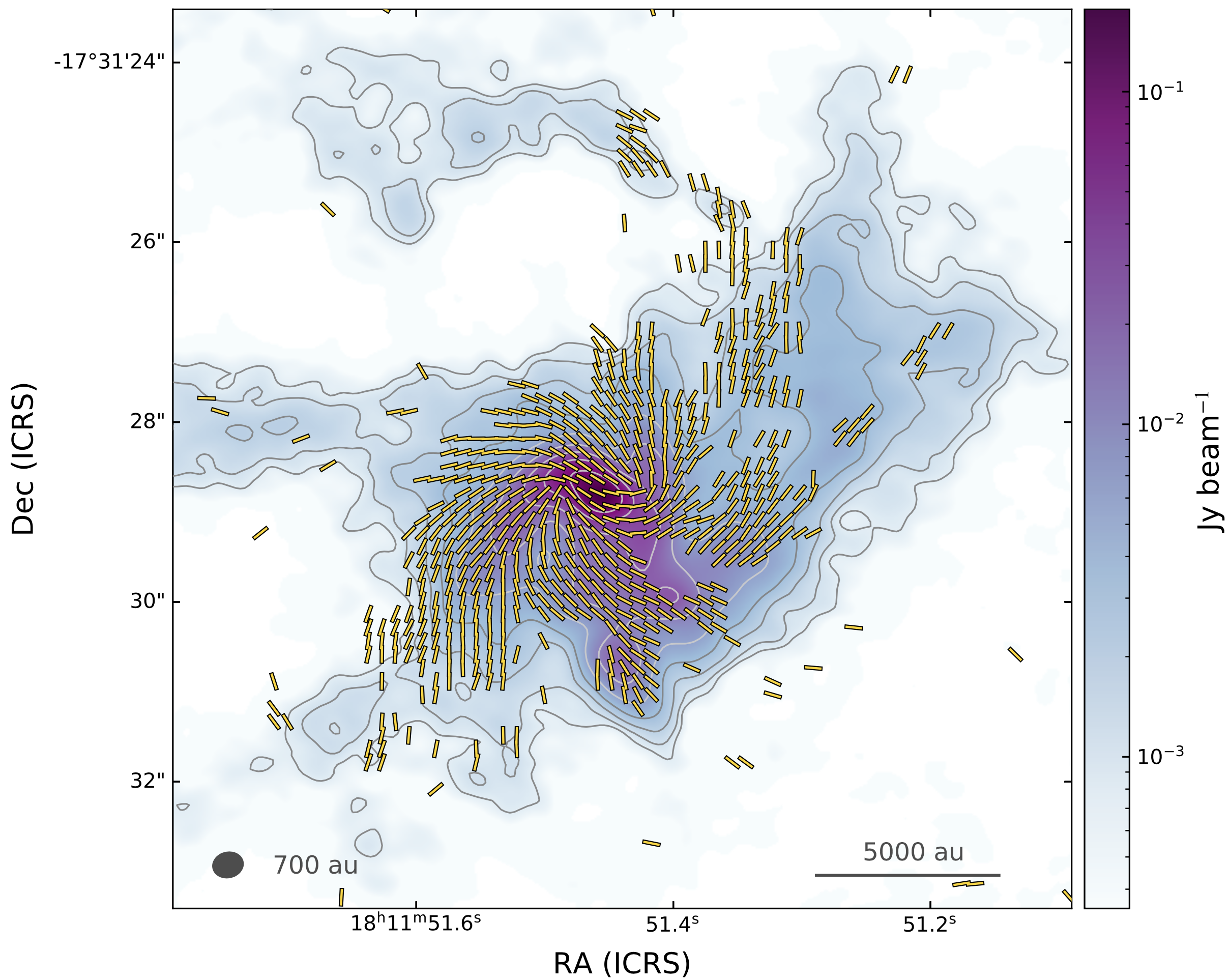}
\caption{ALMA 1.2 mm dust continuum emission (color scale and contours) toward IRAS 18089$-$1732 with overlaid magnetic field vectors, which correspond to the dust polarization vectors rotated by 90 deg. Yellow line segments representing the magnetic field orientation are plotted above the 3$\sigma$ level, with $\sigma=31.4$ $\mu$Jy beam$^{-1}$, and have an arbitrary length. Contours correspond to the dust continuum emission in steps of 4, 6, 10, 18, 34, 66, 130, 258, 514 times the $\sigma$ (rms) value of 175 $\mu$Jy beam$^{-1}$. 
Spatial resolution of 700 au ($0.3''$) is shown on the bottom left. Scale bar is shown on the bottom, right side of the panel.
\label{Bfield}}
\end{figure*}

ALMA polarization observations (Project ID: 2017.1.00101.S; PI: Sanhueza) of IRAS 18089$-$1732 were taken on September 25, 2018. A total of 47 antennas of the 12 m array were used, covering baselines from 15 to 1400 m that resulted in an angular resolution of $\sim$$0.3''$. The data set consists of full polarization observations in band 6 ($\sim$250.486 GHz; 1.2 mm). The correlator setup includes three spectral windows of width 1875 MHz, with a spectral resolution of 1.95 MHz ($\sim$2.4 \kms), and two spectral windows of width 234 MHz, with a spectral resolution of 0.488 MHz (0.56 \kms). 

Linearly polarized dust continuum emission is detected in the inner $\sim$8$''$ of the observed field, the inner one-third of the primary beam ($24''$), with polarization angles having less than 1\% errors. We note, however, that polarization angles on angular scales up to the width of the primary beam have only few percent errors \citep{Hull20}. Line contamination was removed from the continuum (Stokes {\it I}) image following the procedure described in \cite{Olguin21}. Stokes {\it I} was self-calibrated in phase and amplitude, while Stokes {\it Q} and {\it U} were only self-calibrated in phase. Self-calibration solutions were then applied to the spectral cubes. 

The continuum imaging was done by independently cleaning each Stokes parameter using the CASA task {\it tclean} with  Briggs weighting and robust parameter of 1. The resulting images have an angular resolution of $0.27''\x0.34''$ and sensitivities of 175 $\mu$Jy beam$^{-1}$ for Stokes {\it I} and 31.4 $\mu$Jy beam$^{-1}$ for both Stokes {\it Q} and {\it U}. The polarized intensity image was debiased following \cite{Vaillancourt06}.  The peak of the polarized dust emission is 1.4 mJy beam$^{-1}$. The mean (median) polarization fraction is 5\% (4\%).

The H$^{13}$CO$^+$ line emission was imaged using the automatic masking procedure {\it yclean} from \cite{Contreras18}. The CASA task {\it tclean} with Briggs weighting and robust parameter of 1 was used, resulting in a noise level of 
3.2 mJy beam$^{-1}$ per 0.56 \kms\ channel. 

 The quasar J1924-2914 was used for the calibration of flux, bandpass, and polarization. The quasar J1832-2039 was used for phase calibration. Data calibration and imaging were performed using CASA 5.1.1 and 5.5.0,  respectively. 

\section{Results} \label{sec:res}

The ALMA observations at 1.2 mm with a spatial resolution of 700~au ($0.3''$)  allow us to observe  the internal structure of IRAS 18089$-$1732 in great detail  (Figure~\ref{Bfield}). 
The dust continuum emission peaks at the position of the previously reported hot core \citep{Beuther04a,Beuther04b,Zapata06}. 
The overall emission is asymmetric, with an extension  toward the north-west of the brightest 1.2 mm peak. In addition, there are spiral-like streamers associated with the central hot core having a counterclockwise orientation (mimicking a whirlpool). The magnetic field projected in the plane of the sky also traces spiral-like features connected to the central hot core, roughly following the dusty spirals and making more evident the whirlpool shape (Figure~\ref{Bfield}).

\begin{figure*}[ht!]
\epsscale{1.17}
\plotone{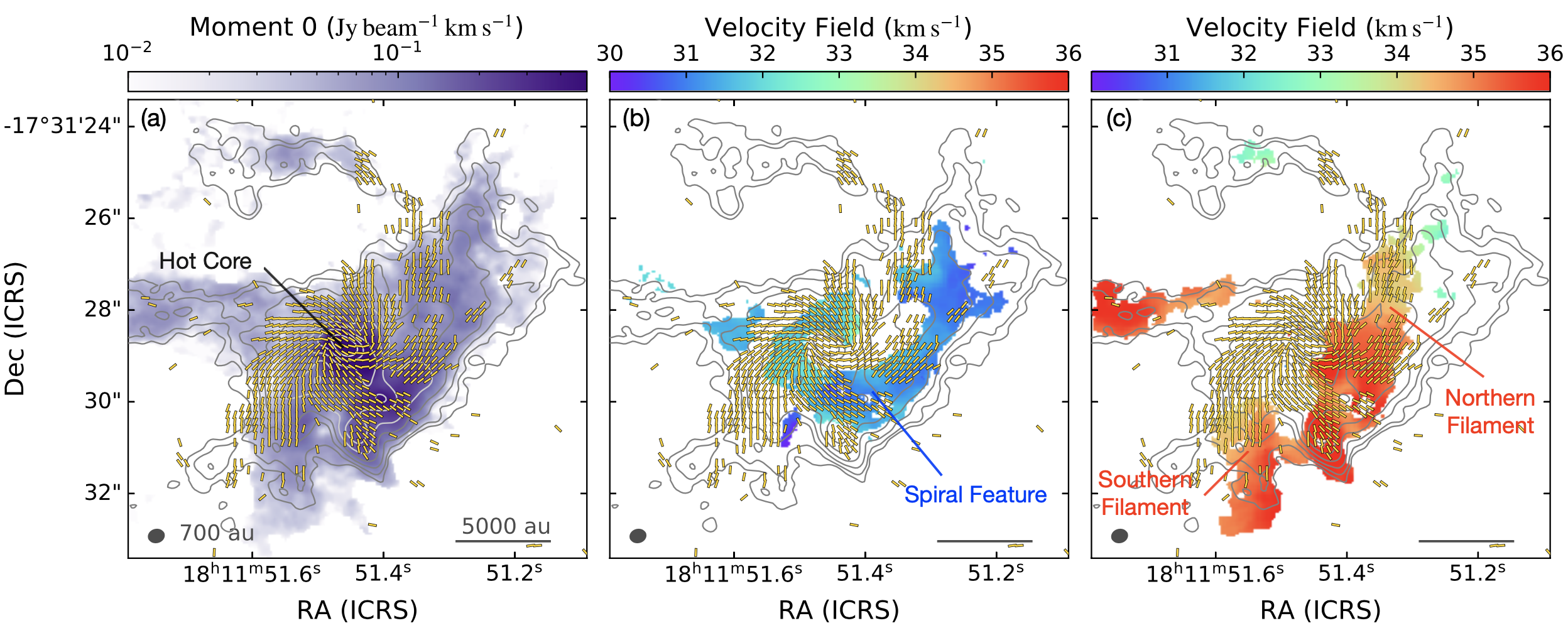}
\caption{H$^{13}$CO$^+$ integrated intensity (moment 0) in panel (a) and velocity fields in panel (b) and (c). Yellow vectors, of an arbitrary length, indicate the inferred magnetic field and gray   contours show the dust continuum emission (same as Figure 1). In (a), the moment 0 includes both velocity components. In (b), the blueshifted component has a spiral-like shape following the magnetic field with a velocity gradient of 17 \kms\ pc$^{-1}$. In (c), the redshifted gas component shows two filaments that follow the magnetic field with velocity gradients of $\sim$54 \kms\ pc$^{-1}$.
\label{line}}
\end{figure*}

Along with the continuum emission, there are myriad  molecular line transitions detected with different excitation conditions (i.e, tracing different temperature and density regimes). Among these, here we focus on 
H$^{13}$CO$^+$ (J = 3--2) and leave the tracers of the inner hot core for a forthcoming work (V. Chen et al. 2021, in prep.). The emission from the H$^{13}$CO$^+$ (J=3-2) line shows a spatial distribution coincident with that of the dust emission, as can be seen in the integrated intensity map (Figure~\ref{line}a). 
This molecule is a good tracer of the relatively cold (E$_u$ = 25 K), dense material in the spiral-like and filamentary structures found in IRAS 18089$-$1732. 
The  H$^{13}$CO$^+$ spectra shows two clear velocity components separated by $\sim$4 km s$^{-1}$ that can be traced continuously from the regions in which they overlap (presenting  double-line profiles) to regions where only a single component is detected. To properly trace the velocity field of both components, a simultaneous Gaussian fitting of both velocity components was performed. Figure~\ref{line}b shows the velocity field traced by the blueshifted gas component (with respect to the systemic cloud velocity $v_{\rm LSR}$ $\approx$ 33.0 \kms), while Figure~\ref{line}c shows the velocity field traced by the redshifted gas component.  Near the center of the hot core, the H$^{13}$CO$^+$ profile becomes complex, exhibiting  extremely broad line widths, absorption features, and blending with hot core lines, precluding a successful Gaussian fitting. 

\begin{figure}[ht!]
\epsscale{1.15}
\plotone{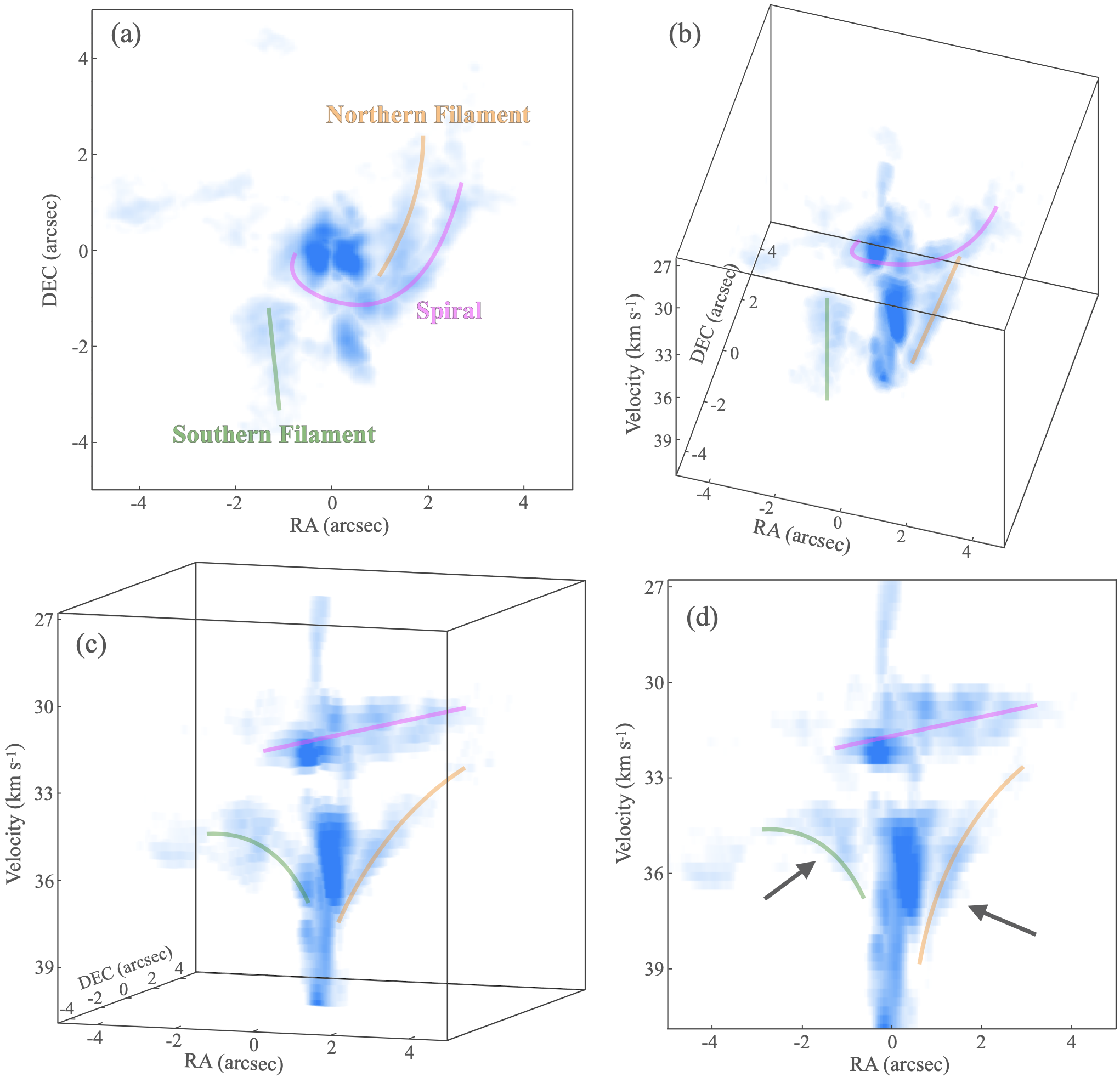}
\caption{Selected views of the position-position-velocity (PPV) distribution of the H$^{13}$CO$^+$ emission. (a) View in position-position (PP) space, comparable to combining Figure~\ref{line}b and~\ref{line}c into a single image. The main structures in IRAS 18089$-$1732 are delineated and labeled. (b) and (c) show the PPV distribution from different angles, with the position of the main features marked. (d) View of the position-velocity (PV) space with the main structures marked. Arrows point to the places where the gas is accelerating and infalling towards the central source. 
\label{3Dimages}}
\end{figure}

A prominent spiral-like feature in the blueshifted gas component  that follows the magnetic field morphology is seen in Figure~\ref{line}b . Although there is significant substructure in the velocity pattern along this  spiral-like filament, it has an overall velocity difference of $\sim$1.4 \kms\ over a length of $7.3''$ (17000 au), resulting in a 17 \kms\ pc$^{-1}$ velocity gradient. The redshifted gas component shows two filaments that coincide with the area emitting polarized emission (Figure~\ref{line}c), one extending to the north and the other one to the south of the central hot core. The northern filament  starts at $3.3''$ (7700 au) from the hot core with blueshifted velocities of $\sim$34.1 \kms\ and then connects to the red-side of the rotating central hot core at $\sim$36.1 \kms\ (velocity gradient of 53.6 \kms\ pc$^{-1}$). The southern  filament extends over $2.9''$ (6800 au) with extreme velocities of $\sim$34.2 and $\sim$36 \kms, resulting in a velocity gradient of 54.6 \kms\ pc$^{-1}$.  In the position-position-velocity (PPV) space displayed in Figure~\ref{3Dimages}, the distribution of the gas and the main structures, spiral and northern/southern filaments, can be seen. Figure~\ref{3Dimages}a simultaneously displays both the blue and redshifted velocity components and is comparable to displaying  Figure~\ref{line}b and~\ref{line}c in a single image. Figure~\ref{3Dimages}b and~\ref{3Dimages}c show the three main structures from different angles. Figure~\ref{3Dimages}d corresponds to a position-velocity (PV) diagram. Figure~\ref{3Dimages}d shows, at the positions indicated by the black arrows, how the velocity increases closer to the central hot core, especially in the southern and northern filaments.  Such an increase in velocity is typical of gas accelerating close to the central object and is a sign of infall \citep{Tobin12}.

\section{Discussion} \label{sec:dis}

\subsection{Dust Continuum Emission}

The physical properties of the central region of IRAS 18089$-$1732 were calculated in the area enclosed by the 10--$\sigma$ contour, which has a flux density of 1.38 Jy. We choose to analyze this area because it includes most of the polarized emission and excludes the most diffuse material and the outlfow cavity \citep[see][]{Beuther04a}. Assuming optically thin dust emission, the total gas mass enclosed in the main dust structure is 75  \Msun\ (Appendix~\ref{sec:dust-app}). The average number density is 1.3 $\times$ 10$^7$ cm$^{-3}$, resulting in a free fall time of 8.4 \x\ $10^3$ yr.

\subsection{Line Emission and Accretion Flows}

The spiral-feature and filaments seen in H$^{13}$CO$^+$ (Figure~\ref{line})  resemble the accretion flows funneling material to disk scales that have been observed (also using high-angular resolution observations) in a few other high-mass star-forming regions \citep[e.g.,][]{Liu15,Maud17,Izquierdo18,Goddi20}.   Following standard procedures, we convert the H$^{13}$CO$^+$ emission tracing the blueshifted spiral-feature and
redshifted filaments into total gas mass (Appendix~\ref{sec:infall-app}).

 The spiral-feature has a gas mass of 2.2 \Msun, while the northern and southern filaments have 1.4 and 0.60 \Msun, respectively. If the velocity gradients observed are produced by material flowing toward the inner hot core, we estimate a combined infall rate of 0.9-2.5\x\ 10$^{-4}$ \Msun\ yr$^{-1}$ for inclination angles between 30$^{\circ}$ and 60$^{\circ}$, larger than the accretion rates through filaments seen in other high-mass star-forming regions at larger scales \citep{Lu18}. 
These infall rates imply that the central disk-like structure can be replenished with 0.70-2.1 \Msun\ per free fall time (8.4 \x\ 10$^3$ yr), thus contributing to the increase in the mass of the central high-mass star.

\subsection{Modeling the Magnetic Field} \label{sec:modeling}

To model the polarization pattern observed in IRAS 18089$-$1732, we used the {\em DustPol} module contained in the ARTIST package \citep{Padovani12} and followed the same procedure described in \cite{Beltran19}. We then perform synthetic ALMA observations of the models and compare the resulting polarization position angles with the observations. {\em DustPol} creates a set of FITS images related to the Stokes parameters ({\it I}, {\it Q}, {\it U}) that is directly used as an input for the CASA simobserve/simanalyze tasks, which use the same antenna configuration of the observing runs.

The model used for the magnetic field configuration is an axially-symmetric singular toroid threaded by an hourglass-shaped poloidal field \citep{Li96,Padovani11} with an added toroidal component to mimic the effect of rotation as in \cite{Padovani13}.  More details on the model are given in Appendix~\ref{sec:model-app}. The main purpose of the modeling is to determine basic properties of the magnetic field (mass-to-magnetic-flux ratio, toroidal-to-poloidal ratio, inclination) to be compared with observations, and to provide the configuration of the mean field needed to perform the Davis-Chandrasekhar-Fermi \citep[DCF;][]{Davis51, CF53} analysis of the polarization angle residuals. 

\begin{figure}[ht!]
\epsscale{1.15}
\plotone{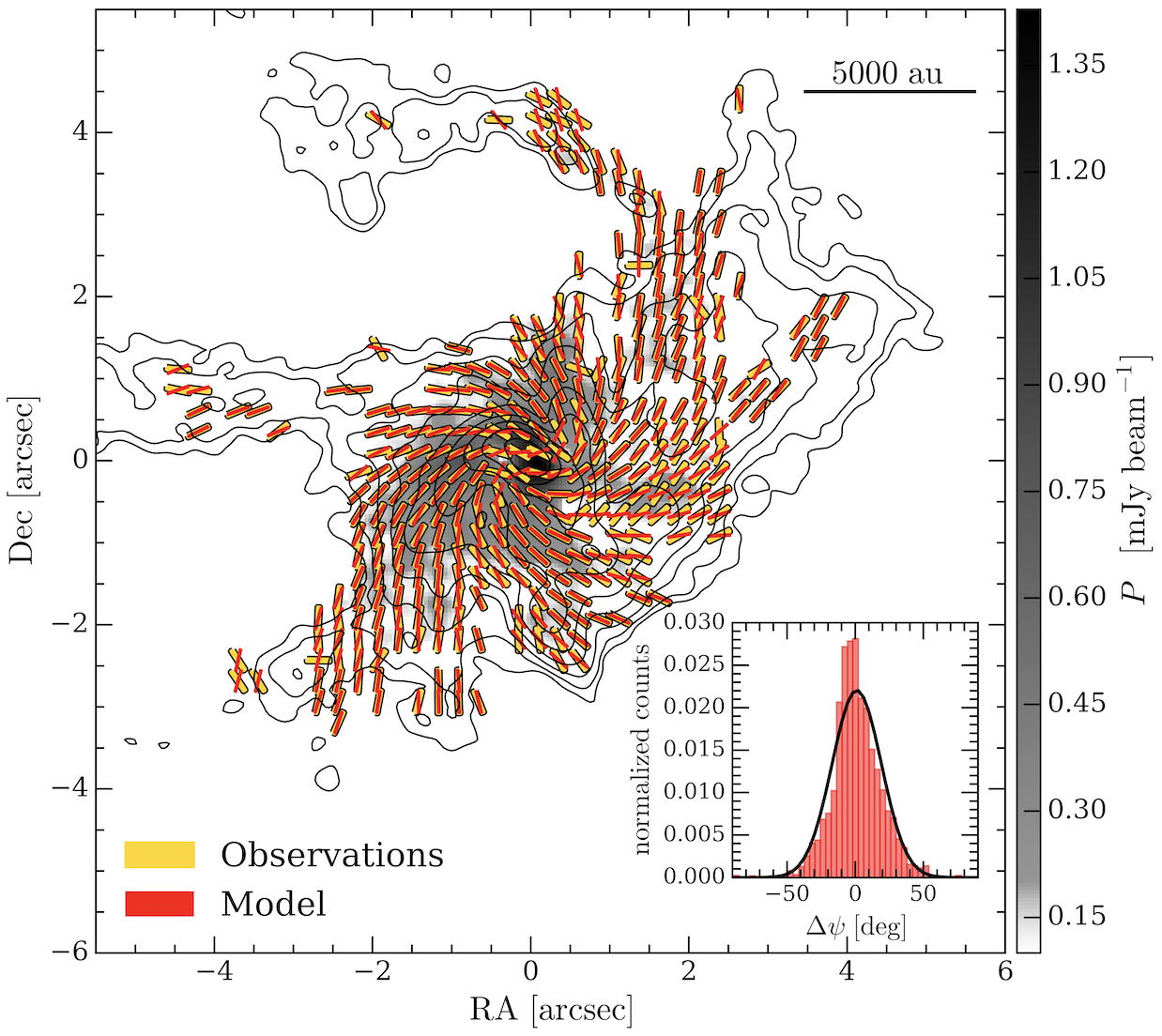}
\caption{Observed polarization angles showing the magnetic field orientation (yellow segments) on top of which are superposed those
obtained from the best-fit model ($\lambda=8.38$, 
$b_0=0.30_{-0.18}^{+0.28}$, and $i=-5_{-15}^{+25}$ deg; red  segments).
The black contours show the dust continuum emission (same as Figure~\ref{Bfield}) and the grey-scale map displays the polarized intensity,
$P$. Scale bar is shown on the top, right side of the panel. 
The inset shows the histogram of the polarization angle residuals $\Delta\psi=\psi_{\rm obs}-\psi_{\rm mod}$ with its 
Gaussian fit (black line).
\label{comparison}}
\end{figure}

We performed a $\chi^2$ test for all the combinations of the parameters $\lambda$, $b_0$, and $i$ and found that the set $\lambda=8.38$, $b_0=0.30_{-0.18}^{+0.28}$, and $i=-5_{-15}^{+25}$~deg gives the lowest reduced $\chi^2$ value ($\bar\chi^2=4.13$). The magnetic field is dominated by a poloidal component, but with an important contribution from the toroidal component, whose magnitude is 30\% of the poloidal component. This significant contribution from the toroidal component indicates that rotation is affecting the magnetic field. Figure \ref{comparison}, shows the comparison between the observed and modeled polarization angles. The excellent agreement is shown quantitatively in the inset of the same figure. The latter illustrates
the distribution of the polarization angle residuals, $\Delta\psi=\psi_{\rm obs}-\psi_{\rm mod}$, defined as the difference between
the observed ($\psi_{\rm obs}$) and modeled ($\psi_{\rm mod}$) polarization angles, whose Gaussian fit gives a mean value of $\langle\Delta\psi\rangle=1.15$ deg and a  standard deviation $\sigma_\psi =18.09$ deg. The latter is needed for the DCF method.  The $\lambda$ value of 8.38 of the best model characterizes the mass inside a flux tube, while observationally the mass is estimated assuming a spherical source. To correct for this, $\lambda$ must be divided by a correction factor of 2.32 \citep{Li96}, leading to an effective $\lambda$ of 3.61. Despite the relative simplicity of the model, this value of 
$\lambda$ is very close to that estimated from observations (see Section~\ref{sec:energy}). It should be stressed that cloud models with 
smaller values of $\lambda$ cannot be completely ruled out. For example, a model with $\lambda=2.66$ also provides an acceptable fit to the data, resulting in $b_0=0.40_{-0.21}^{+0.50}$ and $i=0_{-22}^{+22}$~deg with a slightly worse value of $\bar\chi^2=4.65$ (for smaller values of $\lambda$, the $\bar\chi^2$ becomes progressively larger). We note that the magnetic field strength determined in the following section would not change adopting the $\lambda=2.66$ model because its $\sigma_\psi=18.52$ deg practically has the same value (18.09 deg) as in the preferred $\lambda=8.38$ model.

\subsection{Analysis of the Energy Balance}\label{sec:energy}

To estimate the magnetic field strength, we employ the DCF method, calculating the dispersion of the difference between the observed and modelled polarization angles (Appendix~\ref{sec:bfield-app}). The estimated  magnetic field strength and Alfv\'en speed are 3.5 mG and 1.26 \kms, respectively. 

Once the magnetic field strength is estimated, we can 
assess the importance of each energy component involved in the star formation process in IRAS 18089$-$1732. To put things into context, previous observations find that at low densities and large scales (few to several pc), the magnetic field tends to be parallel to filamentary gas structures \citep[e.g.,][]{Clark14,Planck16}, which in some cases has been  interpreted as material being funneled by the magnetic field toward the cloud center \citep{Palmeirim13}. On the other hand, at higher densities and smaller scales ($\sim$0.5 - 1 pc), the magnetic field is typically found  perpendicular to the gas structures \citep[e.g.,][]{Zhang14,Planck16,Soler17,Liu18,Juvela18,Fissel19,Soam19}. At scales of a few thousand astronomical units (a few times 0.001 pc), however, the relationship between field direction and gas structures is much more uncertain. In IRAS 18089$-$1732, the magnetic field morphology has a spiral pattern at scales from 0.1 pc down to 0.003 pc. The magnetic field vectors tend to follow the spiral-like features and filaments seen in both the continuum as well as in the molecular line  emission. As suggested by the model, the twisted morphology of the magnetic field seems to be produced by gas rotating around the hot core. 

To confirm if gravity has overwhelmed the support that the magnetic field can provide against collapse, we calculate the mass-to-magnetic-flux ratio ($\lambda$; Appendix~\ref{sec:model-app}~and~\ref{sec:bfield-app}). From observations, we obtain a $\lambda$ of 3.2 (consistent with the model value mentioned above after correcting for geometry), indicating that the core is supercritical and likely  collapsing. Indeed, the magnetic to gravitational energy ratio is 0.11. Turbulence is an additional source of energy that can oppose gravity and, in low-mass star-forming regions, turbulence  has been found to be more dynamically important than the magnetic field \citep{Hull17}. Their relative importance can be assessed by the turbulent to magnetic energy ratio, $\beta_t$. We obtain a $\beta_t$ of 1.1, which indicates that turbulence and the magnetic field play a comparable role in supporting the IRAS 18089$-$1732 core against collapse. With an excellent agreement between the observations and the model seen in Figure~\ref{comparison}, the important contribution from the toroidal component of the magnetic field reveals the effects of rotation, suggesting that the rotational and magnetic field energy should be  comparable. We estimate the rotational energy of the system (Appendix~\ref{sec:bfield-app}) and find that it is also 
overcome by the gravitational energy and it is only slightly weaker than the magnetic field energy. To assess the relative importance of all energies at play, the virial analysis is frequently performed. In its basic form, the virial analysis includes only gravity and turbulence, and relies on the determination of the virial parameter ($\alpha_{\rm vir}$). We computed a virial parameter of 0.24 (unity indicates equilibrium), which indicates
collapse. The overwhelming importance of gravity with respect to the other energies becomes more evident once we include the magnetic field and rotation. 
After adding the magnetic field and then rotation, we obtain
virial parameters of 0.35 ($\alpha_{\rm vir,B}$) and 0.57 ($\alpha_{\rm vir,B,rot}$; see derivation in Appendix~\ref{sec:bfield-app}), respectively. These calculations of the virial parameter assume a uniform density profile. If a centrally peaked density profile is used, virial parameters are 50-60\% smaller: 0.14 ($\alpha_{\rm vir}$), 0.21 ($\alpha_{\rm vir,B}$), and 0.28 ($\alpha_{\rm vir,B,rot}$), diminishing the importance of other energies with respect to gravity. 

Overall, at the scales we observe in IRAS 18089$-$1732, the magnetic field morphology seems to be dictated by dynamical processes such as inflow and rotation, thus indicating a diminished importance of the magnetic field at the current evolutionary stage. The analysis of the energies in play supports that gravity has taken the dominant role in the immediate vicinity ($\sim$1000 au scales) of a high-mass star in formation, and the magnetic field importance is only comparable to turbulence and rotation.

\section{Conclusions} \label{sec:conclu}

ALMA observations of the high-mass star-forming region IRAS~18089$-$1732 have revealed that the dense molecular envelope surrounding the high-mass star has a complex spiral pattern at the 0.003--0.1 pc scales. This spiral-like morphology is seen in the gas and dust (traced by the H$^{13}$CO$^+$ (J=3-2) line and the 1.2 mm continuum emission, respectively), as well as in the magnetic field (traced by the linearly polarized 1.2~mm dust emission). At the observed size scales, gravitational infall clearly dominates over the support from the magnetic field, turbulence, and rotation, resulting in the feeding of the inner dense, hot circumstellar disk-like structure with a high accretion rate of 0.9-2.5\x\ 10$^{-4}$ \Msun\ yr$^{-1}$. 
We show that high-mass star formation can occur in weakly magnetized environments and that gravity is shaping the immediate surrounding around the high-mass star. The spiral magnetic field indicates that angular momentum is high enough to twist the field lines, as supported by the model and the energy analysis. With these observations and consistent with previous works in other high-mass star-forming regions \citep[e.g.,][]{Koch14,Koch18}, we suggest that the importance of the magnetic field in the process of high-mass star formation depends on the size scales traced and the evolutionary stage of the observed region.

\begin{acknowledgments}
P.S. is thankful of the anonymous referee. P.S. was partially supported by a Grant-in-Aid for Scientific Research (KAKENHI Number 18H01259) of the Japan Society for the Promotion of Science (JSPS). J.M.G. acknowledges the support of the Spanish grant AYA2017-84390-C2-R (AEI/FEDER, UE). C.L.H.H. acknowledges the support of the NAOJ Fellowship and JSPS KAKENHI grants 18K13586 and 20K14527. 
JMJ's research was conducted in part at the SOFIA Science Center, which is operated by the Universities Space Research Association under contract NNA17BF53C with the National Aeronautics and Space Administration. K.T. was supported by JSPS KAKENHI Grant Number 20H05645. Data analysis was in part carried out on the Multi-wavelength Data Analysis System operated by the Astronomy Data Center (ADC), National Astronomical Observatory of Japan.
This paper makes use of the following ALMA data: ADS/JAO.ALMA\#2017.1.00101.S. ALMA is a partnership of ESO (representing its member states), NSF (USA) and NINS (Japan), together with NRC (Canada), MOST and ASIAA (Taiwan), and KASI (Republic of Korea), in cooperation with the Republic of Chile. The Joint ALMA Observatory is operated by ESO, AUI/NRAO and NAOJ.
\end{acknowledgments}

%

\vspace{5mm}
\facilities{ALMA}


\software{CASA \citep[v5.1.1, 5.5; ][]{McMullin07}}



\appendix

\section{Properties from Dust Continuum Emission} \label{sec:dust-app}

The total gas mass can be calculated from the dust emission, in the optically thin limit, as 
\begin{equation}
    M = \mathbb{R}~\frac{F_\nu D^2}{\kappa_\nu B_\nu (T)}~,
\label{eqn-dust-mass}
\end{equation}
where $F_\nu$ is the source flux density, $\mathbb{R}$ is the gas-to-dust mass ratio, 
$D$ is the distance to the source, $\kappa_\nu$ is the dust opacity per gram of dust, and $B_\nu$ is
 the Planck function at the dust temperature $T$. Assuming a dust-to-gas mass ratio of 100, dust opacity of 1.03 cm$^2$ g$^{-1}$ \citep[interpolated to 1.2 mm assuming $\beta$ = 1.6;][]{OH94}, and a temperature of 30 K \citep{Lu14}, the  computed total mass is 75 \Msun. The number density, $n$(H$_2$) = $M$/(Volume \x\ $\mu_{\rm H_2}m_{\rm H}$) with $\mu_{\rm H_2}$ the molecular weight per hydrogen molecule and $m_{\rm H}$ the hydrogen mass, and the surface density, $\Sigma = M/(\pi r^2)$,  can be calculated assuming a spherical core.  We define an effective radius ($r_{\rm eff}$) determined by the area ($A$) emitting above 10$\sigma$ as $r_{\rm eff} = (A/\pi)^{1/2}$, resulting in $r_{\rm eff}=2.36''$ (equal to 0.027 pc or 5500 au). Assuming $\mu_{\rm H_2}=2.8$, we obtain a $n({\rm H}_2)$ of 1.3 \x\ $10^7$~cm$^{-3}$ and $\Sigma=6.9$ g cm$^{-2}$, values characteristic of cores forming high-mass stars. The free fall time is determined from 
 \begin{equation}
 t_{\rm ff} = \sqrt{\frac{3 \pi}{32 G \rho}}~,
 \end{equation}
  where $G$ is the gravitational constant and $\rho$ is the mass density, $n({\rm H}_2$) \x\ $\mu_{\rm H_2}m_{\rm H}$. With $\rho$ equal to 6.3 \x\ 10$^{-17}$ g cm$^{-3}$, the free fall time is 8.4 \x\ 10$^3$ yr.

\section{Properties from Line Emission} \label{sec:infall-app}

We estimated the total gas mass in the spiral-like feature and filaments from the H$^{13}$CO$^+$ emission following a standard procedure \citep{Sanhueza12} as follows. First, the column density for a linear, rigid rotor in the optically thin regime, assuming a filling factor of unity, can be calculated from
\begin{equation}
N=\frac{3k}{8\pi^3B_{\rm rot}\mu^2}\frac{(T_{\rm ex}+hB_{\rm rot}/3k_{\rm B})}{(J+1)}\frac{\exp(E_{_{J}}/kT_{\rm ex})}{[1-\exp(-h\nu/kT_{\rm ex})]}\frac{1}{[J(T_{\rm ex}) - J(T_{\rm bg})]}\int T_{\rm b}\,dv~,~~~~~~  
\label{eqn-den-colum-thin}
\end{equation}
where $k_{\rm B}$ is the Boltzmann constant, $h$ is the Planck constant, $T_{\rm ex}$ is the excitation temperature \citep[30 K;][]{Lu14}, $\nu$ is the transition frequency (260.255339 GHz), $\mu$ is the permanent dipole moment of the molecule (3.89 D), $J$ is the rotational quantum number of the lower state, $E_J=hB_{\rm rot}J(J+1)$ is the energy in the level $J$, $B_{\rm rot}$ is the rotational constant of the molecule (43.377302 GHz), $T_{\rm b}$ is the brightness temperature, $T_{\rm bg}$ the background temperature, and $J(T)$ is defined as 
 \begin{equation}
J(T) = \frac{h\nu}{k}\frac{1}{e^{h\nu/kT} - 1}~. 
\label{eqn-J}
\end{equation}

The column density is then converted into mass using 
\begin{equation}
M = \left[\frac{\rm H^{13}CO^+}{\rm H_2}\right]^{-1} \mu_{\rm H_2} A \sum N(\rm {H^{13}CO^+})~,
\label{eqn-mass-thin}
\end{equation}
where $A$ is the size of the emitting area in an individual
position (pixel of 0.05$''$), [H$^{13}$CO$^+$/H$_2$] is the H$^{13}$CO$^+$ to molecular hydrogen abundance ratio, and the sum is over all the observed positions. We have assumed an [H$^{13}$CO$^+$/H$_2$] abundance ratio typical of high-mass star-forming regions equal to 1.28 \x\ 10$^{-10}$  \citep{Hoq13}. 

\section{Magnetic Field Modeling} \label{sec:model-app}

The idealized model used represents a cloud in magnetostatic equilibrium with density and magnetic field strength decreasing with radius as $r^{-2}$ and $r^{-1}$, respectively.  The field lines are spirals wrapping around nested hourglass-shaped magnetic flux tubes, and have a kink in the equatorial plane where the toroidal component changes sign. 
The parameters of the model are: (i) the mass-to-magnetic-flux ratio normalized to its critical value, $\lambda$, given by 
\begin{equation}
    \lambda = \frac{(M/\Phi_B)}{(M/\Phi_B)_{\rm cr}} = 2 \pi G^{1/2}\frac{M}{\Phi_B}~,
\label{lambda}    
\end{equation}
where $G$ is the gravitational constant, $M$ is the core mass, and $\Phi_B$ is the magnetic flux; (ii) the ratio of the strength of the toroidal and poloidal magnetic field components in the midplane, 
$b_0$; 
and (iii) the inclination of the toroid with respect to the plane of the sky, $i$ ($i=0$ for face-on). The latter is assumed positive if the magnetic field in the north side is directed toward the observer. In principle the orientation of the  projection of the magnetic axis on the plane of the sky, $\varphi$, should also be considered as a further parameter,
but as we will show, the best-fit model indicates that the equatorial plane of the cloud is practically in the plane of the sky ($i=-5$~deg), so that the model turns out to be insensitive to $\varphi$.

We considered four values for $\lambda$. The extreme cases include a nearly spherical density profile  with a weak magnetic field ($\lambda=16.2$) and a flat density profile with a strong magnetic field ($\lambda=1.63$). The two intermediate cases correspond to $\lambda=2.66$ and 8.38. 
For the magnetic field configuration, we have considered the range of $b_0$ from the pure poloidal case ($b_0=0$) to the case where the strength of the toroidal component is twice the poloidal one ($b_0=2$). Finally, we have taken inclinations with respect to the plane of the sky between $-90$ and 90 deg.

\begin{figure*}[ht!]
\epsscale{1.0}
\plotone{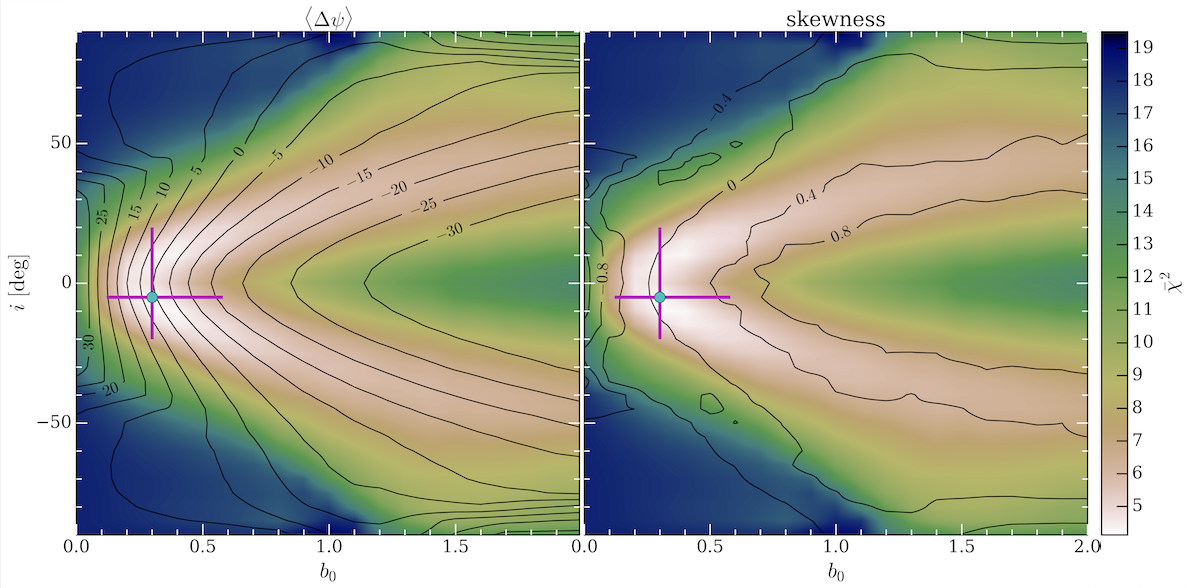}
\caption{Map of the $\bar\chi^2$ distribution of residuals as a function of the inclination, $i$, and the toroidal-to-poloidal ratio, $b_0$, for the model with $\lambda=8.38$. The black isocontours show 
the average value of the polarization angle residuals $\Delta\psi=\psi_{\rm obs}-\psi_{\rm mod}$ (left panel) and the skewness (also known as moment 3) of their distribution (right panel). The solid cyan circle and the magenta lines show the best-fit values of $b_0$ and $i$ and their uncertainties, respectively.
\label{comparison2}}
\end{figure*}

The lowest  $\chi^2$ value, 4.13, is found for the set $\lambda=8.38$, $b_0=0.30_{-0.18}^{+0.28}$, and $i=-5_{-15}^{+25}$~deg \citep[uncertainties on $b_0$ and $i$ have been estimated using the method of][]{Lampton76}. The best fit model can be seen in Figure \ref{comparison}. For completeness, Figure~\ref{comparison2} shows the map of the $\bar\chi^2$ distribution as a function of the explored range of $b_0$ and $i$. Superimposed on this map are the isocontours of the average value of the polarization angle residuals, $\langle\Delta\psi\rangle$ (left panel), and the skewness (also known as moment 3) parameter of their distribution (right panel). The mass-to-magnetic-flux ratio that corresponds to the minimum $\bar\chi^2$ also gives the distribution of $\Delta\psi$ with the smallest degree of skewness (equal to 0.12, namely the distribution is only
slightly positively skewed, which means that the peak of the distribution is shifted towards negative values as seen in the inset of Figure~\ref{comparison}).

\section{Magnetic Field Properties and Energy Balance} \label{sec:bfield-app}

The magnitude of the magnetic field was estimated using the DCF method, following the procedure of \cite{Beltran19} 
\begin{equation}
    B_{\rm pos} = \xi \frac{\sigma_{\rm los}}{\delta \psi_{\rm int}} \sqrt{4\pi\rho}~, 
\end{equation}
where $\xi = 0.5$ is a correction factor derived from turbulent cloud simulations \citep{Ostriker01} and $\sigma_{\rm los}$ is the line-of-sight velocity dispersion. The DCF method assumes that $\sigma_{\rm los}$ is from turbulence, and indeed, the observed velocity dispersion is 2.6 times the thermal line width. 
The average velocity dispersion  ($\sigma_{\rm los}$) for both H$^{13}$CO$^+$ velocity components above the 10$\sigma$ area, in the continuum image, is 0.75 km s$^{-1}$ (both blueshifted and redshifted components have practically the same line width). The density used in the estimation of the magnetic field strength is 6.3 \x\ 10$^{-17}$ g cm$^{-3}$.  $\delta \psi_{\rm int}$ is the intrinsic angle dispersion given by  $\delta \psi_{\rm int}$ = $(\sigma_\psi^2 - \delta \psi_{\rm obs}^2)^{1/2}$. $\sigma_\psi$ of 18.1 deg is the standard deviation of the polarization angle residuals obtained from the magnetic field modeling (see inset in Figure~\ref{comparison}, top) and $\delta \psi_{\rm obs}$ of 6.0 deg is the mean angle uncertainty in the whole debiased polarization angle image. The derived $\delta \psi_{\rm int}$ of 17.1 deg results in a magnetic field strength of B$_{\rm pos}$ = 3.5 mG. The Alfv\'en speed, given by $\sigma_A = B/\sqrt{4\pi\rho}$, is 1.26 \kms. 

We calculate the mass-to-magnetic-flux ratio normalized to the critical value, $\lambda$, using equation (\ref{lambda}) and the fact that  
$\Phi_B = B \pi R^2$. Adopting $R=r_{\rm eff}$, we obtain a value of $\lambda$ equal to 3.2. 

The ratio of the turbulent to magnetic energy, $\beta_t$, is given by 
\begin{equation}
    \beta_t = 3 \left(\frac{\sigma_{\rm los}}{\sigma_A}\right)^2~.
\end{equation}
The calculated $\beta_t$ is 1.07, which indicates that turbulence and the magnetic field play a comparable role in the energy budget of the system.  

We assess the importance of the rotational energy with respect to gravity and the magnetic field as follows. First, we take the ratio between the gravitational energy, $E_{\rm G}$,
\begin{equation}
E_{\rm G} = -\frac{GM^2}{R}\left(\frac{3 - \alpha}{5 - 2\alpha} \right)~,
\end{equation}
and the rotational energy, $E_{\rm rot}$, 
\begin{equation}
E_{\rm rot} = \frac{1}{2} I \Omega^2 = \frac{1}{3}Mv_{\rm rot}^2\left(\frac{3 - \alpha}{5 - \alpha} \right)~,
\end{equation}
where the moment of inertia ($I$) of a sphere has been assumed, the angular velocity is given by $\Omega = v_{\rm rot}/R$, $v_{\rm rot}$ is the rotational velocity, and $\alpha$ corresponds to the exponent in a density profile of the form $\rho(R) \propto  R^{-\alpha}$ \citep[where $\alpha$ = 0 implies a uniform density profile and $\alpha$ = 2 a centrally peaked density profile;][]{Belloche13}. The maximum velocity gradient observed is of 2 \kms. If we assume as upper limit that this gradient is purely produced by rotation, we obtain a $E_{\rm rot}/E_{\rm G}$ ratio of 0.11 for a uniform density profile ($\alpha$ = 0) and 0.037 for a centrally peak density profile ($\alpha$ = 2).

To evaluate the relative importance between rotation and the magnetic field, we follow two approaches: the rotation to magnetic field energy ratio and the prescription from \cite{Machida05}. 
First, the energy in the magnetic field ($E_{\rm B}$) is given by
\begin{equation}
E_{\rm B} = \frac{1}{8\pi}B^2V = \frac{1}{6}B^2R^3~,
\end{equation}
where $V$ is the volume, assumed here to be a sphere. Taking into account the same consideration from above (velocity gradient entirely produced by rotation), the $E _{\rm rot}/E_{B}$ ratio is 1.0 for a uniform density distribution and 0.56 for a centrally peaked density distribution. These results suggest that the magnetic field is slightly more important than rotation, depending on the exact density profile.  

Second, for assessing whether the magnetic field dominates over rotation during the collapse of clouds, \cite{Machida05} suggest that the ratio of the angular velocity to the magnetic flux density can be used. According to the following relation, if the observed ratio is
\begin{equation}
\left(\frac{\Omega}{B}\right)_{\rm obs}>0.39\frac{\sqrt{G}}{\sigma_{\rm th}} = 1.69 \times 10^{-7} \left(\frac{\sigma_{\rm th}}{0.19~{\rm km ~s}^{-1}}\right)^{-1} \rm{yr}^{-1} ~\mu \rm{G}^{-1}~,
\label{eq-omega_B}
\end{equation}
rotation dominates over the magnetic field, otherwise the magnetic field dominates over rotation. Evaluating the thermal velocity dispersion (isothermal sound speed), $\sigma_{\rm th}=\left( \frac{k_B T}{\mu m_H}\right)^{1/2}$, for a temperature of 30 K and using the mean molecular
weight per free particle $\mu=2.37$, we obtain the right hand side of Equation~\ref{eq-omega_B} equal to 9.8 \x\ 10$^{-8}$ yr$^{-1}$ $\mu$G$^{-1}$. Evaluating the observed ratio, we obtain 2.2 \x\ 10$^{-8}$ yr$^{-1}$ $\mu$G$^{-1}$. 
The observed ratio is lower 
than the right-hand side of Equation~\ref{eq-omega_B} by a factor 4.5, suggesting that from this analysis the magnetic field dominates over rotation. Based on both methods employed to assess the importance of rotation with respect to the magnetic field, we conclude that the magnetic field is slightly more important than rotation.

The dynamical state of cores is generally evaluated by using the virial theorem. The ratio between the virial mass, $M_{\rm vir}$, and the total mass defines the virial parameter, $\alpha_{\rm vir}$. A virial parameter of unity implies equilibrium, $\alpha_{\rm vir} < 1$ implies gravitational collapse, $\alpha_{\rm vir} > 1$ means the core is expanding and it will  disperse. In its simplest form, the virial analysis only includes  gravity and the kinetic energy (turbulence and thermal energy) as follow
\begin{equation}
    \alpha_{\rm vir}=\frac{M_{\rm vir}}{M}=3\left(\frac{5-2\alpha}{3-\alpha}\right)\frac{R\sigma_{\rm los}^2}{GM}~,
\label{eq-virial_1}
\end{equation}
resulting in $\alpha_{\rm vir}=0.24$ for a uniform density profile ($\alpha=0$) and 0.14 for a centrally peaked density profile ($\alpha=2$). Both low $\alpha_{\rm vir}$ values, in the absence of other energies, indicate that the core is collapsing dominated by gravity. 

The virial parameter considering the magnetic field, is written as follow \citep[e.g.,][]{Sanhueza17}
\begin{equation}
    \alpha_{\rm vir,B}=\frac{M_{\rm vir}}{M}=3\left(\frac{5-2\alpha}{3-\alpha}\right)\frac{R}{GM} \left(\sigma_{\rm los}^2 + \frac{\sigma_A^2}{6}\right),
\label{eq-virial_2}
\end{equation}
which results in $\alpha_{\rm vir,B}$ of 0.35 for a uniform density profile and 0.21 for a centrally peaked density profile. Including the magnetic field, gravity continue to be dominant and the core is collapsing.  

We derive the contribution of rotation in the virial equation as below. The virial theorem can be written as
\begin{eqnarray}
\frac{1}{2}\frac{d^2I}{dt^2} & = & 2E_{\rm k} + E_{\rm G} + E_{\rm B}~,\\
& = & 2(E_{\rm n-rot} + E_{\rm rot}) + E_{\rm G} + E_{\rm B}~,
\label{eq-virial}
\end{eqnarray}
in which $E_{\rm k}$ is the kinetic energy that can be separated in a rotational part ($E_{\rm rot}$) and non-rotational part \citep[$E_{\rm n-rot;}$][]{McKee92}. The non-rotational part includes thermal and turbulent energies as follow
\begin{equation}
E_{\rm n-rot} = \frac{3}{2}M\sigma_{\rm los}^2~,
\end{equation}
where the observed velocity dispersion is $\sigma_{\rm los}^2 = \sigma_{\rm th}^2 + \sigma_{\rm tur}^2$ and $\sigma_{\rm tur}$ is the turbulent component. $E_{\rm n-rot}$ is frequently assumed as $E_{\rm k}$ when rotation is neglected.  In virial equilibrium, the moment of inertia does not vary over time and the left-hand side of Equation~\ref{eq-virial} becomes zero. Solving Equation~\ref{eq-virial} for the mass, one can find the virial mass and the virial parameter can be calculated. If rotation and magnetic energies are ignored, Equation~\ref{eq-virial_1} is recovered. If only rotation is ignored, Equation~\ref{eq-virial_2} is found. Solving for all energies in play, we obtain
\begin{equation}
    \alpha_{\rm vir,B,rot}=3\left(\frac{5-2\alpha}{3-\alpha}\right)\frac{R}{GM} \left(\sigma_{\rm los}^2 + \frac{\sigma_A^2}{6} + \frac{2}{9}\left(\frac{3-\alpha}{5-\alpha}\right)v_{\rm rot}^2\right).
\label{eq-virial_3} 
\end{equation}
For a uniform density profile and for a centrally peaked density distribution, we obtain a virial parameter of 0.57 and 0.28, respectively. This implies that at the scales we probe with our ALMA observations, even when both the magnetic field and rotation are considered, gravity is still the dominant dynamical force (independently of the density profile assumed).


\bibliography{references}{}

\begin{thebibliography}{}
\expandafter\ifx\csname natexlab\endcsname\relax\def\natexlab#1{#1}\fi
\providecommand{\url}[1]{\href{#1}{#1}}
\providecommand{\dodoi}[1]{doi:~\href{http://doi.org/#1}{\nolinkurl{#1}}}
\providecommand{\doeprint}[1]{\href{http://ascl.net/#1}{\nolinkurl{http://ascl.net/#1}}}
\providecommand{\doarXiv}[1]{\href{https://arxiv.org/abs/#1}{\nolinkurl{https://arxiv.org/abs/#1}}}

\bibitem[{{Belloche}(2013)}]{Belloche13}
{Belloche}, A. 2013, in EAS Publications Series, Vol.~62, EAS Publications
  Series, ed. P.~{Hennebelle} \& C.~{Charbonnel}, 25--66

\bibitem[{{Beltr{\'a}n} {et~al.}(2019){Beltr{\'a}n}, {Padovani}, {Girart},
  {Galli}, {Cesaroni}, {Paladino}, {Anglada}, {Estalella}, {Osorio}, {Rao},
  {S{\'a}nchez-Monge}, \& {Zhang}}]{Beltran19}
{Beltr{\'a}n}, M.~T., {Padovani}, M., {Girart}, J.~M., {et~al.} 2019, \aap,
  630, A54, \dodoi{10.1051/0004-6361/201935701}

\bibitem[{{Beuther} {et~al.}(2010){Beuther}, {Vlemmings}, {Rao}, \& {van der
  Tak}}]{Beuther10}
{Beuther}, H., {Vlemmings}, W.~H.~T., {Rao}, R., \& {van der Tak}, F.~F.~S.
  2010, \apjl, 724, L113, \dodoi{10.1088/2041-8205/724/1/L113}

\bibitem[{{Beuther} {et~al.}(2004{\natexlab{a}}){Beuther}, {Hunter}, {Zhang},
  {Sridharan}, {Zhao}, {Sollins}, {Ho}, {Ohashi}, {Su}, {Lim}, \&
  {Liu}}]{Beuther04a}
{Beuther}, H., {Hunter}, T.~R., {Zhang}, Q., {et~al.} 2004{\natexlab{a}},
  \apjl, 616, L23, \dodoi{10.1086/383570}

\bibitem[{{Beuther} {et~al.}(2004{\natexlab{b}}){Beuther}, {Zhang}, {Hunter},
  {Sridharan}, {Zhao}, {Sollins}, {Ho}, {Liu}, {Ohashi}, {Su}, \&
  {Lim}}]{Beuther04b}
{Beuther}, H., {Zhang}, Q., {Hunter}, T.~R., {et~al.} 2004{\natexlab{b}},
  \apjl, 616, L19, \dodoi{10.1086/422500}

\bibitem[{{Beuther} {et~al.}(2020){Beuther}, {Soler}, {Linz}, {Henning},
  {Gieser}, {Kuiper}, {Vlemmings}, {Hennebelle}, {Feng}, {Smith}, \&
  {Ahmadi}}]{Beuther20}
{Beuther}, H., {Soler}, J.~D., {Linz}, H., {et~al.} 2020, \apj, 904, 168,
  \dodoi{10.3847/1538-4357/abc019}

\bibitem[{{Chandrasekhar} \& {Fermi}(1953)}]{CF53}
{Chandrasekhar}, S., \& {Fermi}, E. 1953, \apj, 118, 116,
  \dodoi{10.1086/145732}

\bibitem[{{Clark} {et~al.}(2014){Clark}, {Peek}, \& {Putman}}]{Clark14}
{Clark}, S.~E., {Peek}, J.~E.~G., \& {Putman}, M.~E. 2014, \apj, 789, 82,
  \dodoi{10.1088/0004-637X/789/1/82}

\bibitem[{{Contreras} {et~al.}(2018){Contreras}, {Sanhueza}, {Jackson},
  {Guzm{\'a}n}, {Longmore}, {Garay}, {Zhang}, {Nguyen-Lu'o'ng}, {Tatematsu},
  {Nakamura}, {Sakai}, {Ohashi}, {Liu}, {Saito}, {Gomez}, {Rathborne}, \&
  {Whitaker}}]{Contreras18}
{Contreras}, Y., {Sanhueza}, P., {Jackson}, J.~M., {et~al.} 2018, \apj, 861,
  14, \dodoi{10.3847/1538-4357/aac2ec}

\bibitem[{{Cortes} {et~al.}(2019){Cortes}, {Hull}, {Girart}, {Orquera-Rojas},
  {Sridharan}, {Li}, {Louvet}, {Cortes}, {Le Gouellec}, {Crutcher}, \&
  {Lai}}]{Cortes19}
{Cortes}, P.~C., {Hull}, C. L.~H., {Girart}, J.~M., {et~al.} 2019, \apj, 884,
  48, \dodoi{10.3847/1538-4357/ab378d}

\bibitem[{{Cortes} {et~al.}(2021){Cortes}, {Le Gouellec}, {Hull}, {Girart},
  {Louvet}, {Fomalont}, {Kameno}, {Moellenbrock}, {Nagai}, {Nakanishi}, \&
  {Villard}}]{Cortes21}
{Cortes}, P.~C., {Le Gouellec}, V. J.~M., {Hull}, C. L.~H., {et~al.} 2021,
  \apj, 907, 94, \dodoi{10.3847/1538-4357/abcafb}

\bibitem[{{Dall'Olio} {et~al.}(2017){Dall'Olio}, {Vlemmings}, {Surcis},
  {Beuther}, {Lankhaar}, {Persson}, {Richards}, \& {Varenius}}]{Olio17}
{Dall'Olio}, D., {Vlemmings}, W.~H.~T., {Surcis}, G., {et~al.} 2017, \aap, 607,
  A111, \dodoi{10.1051/0004-6361/201731297}

\bibitem[{{Davis}(1951)}]{Davis51}
{Davis}, L. 1951, Physical Review, 81, 890, \dodoi{10.1103/PhysRev.81.890.2}

\bibitem[{{Fern{\'a}ndez-L{\'o}pez} {et~al.}(2021){Fern{\'a}ndez-L{\'o}pez},
  {Sanhueza}, {Zapata}, {Stephens}, {Hull}, {Zhang}, {Girart}, {Koch},
  {Cort{\'e}s}, {Silva}, {Tatematsu}, {Nakamura}, {Guzm{\'a}n}, {Nguyen Luong},
  {Guzm{\'a}n Ccolque}, {Tang}, \& {Chen}}]{Fernandez21}
{Fern{\'a}ndez-L{\'o}pez}, M., {Sanhueza}, P., {Zapata}, L.~A., {et~al.} 2021,
  arXiv e-prints, arXiv:2104.03331.
\newblock \doarXiv{2104.03331}

\bibitem[{{Fissel} {et~al.}(2019){Fissel}, {Ade}, {Angil{\`e}}, {Ashton},
  {Benton}, {Chen}, {Cunningham}, {Devlin}, {Dober}, {Friesen}, {Fukui},
  {Galitzki}, {Gandilo}, {Goodman}, {Green}, {Jones}, {Klein}, {King},
  {Korotkov}, {Li}, {Lowe}, {Martin}, {Matthews}, {Moncelsi}, {Nakamura},
  {Netterfield}, {Newmark}, {Novak}, {Pascale}, {Poidevin}, {Santos}, {Savini},
  {Scott}, {Shariff}, {Soler}, {Thomas}, {Tucker}, {Tucker}, {Ward-Thompson},
  \& {Zucker}}]{Fissel19}
{Fissel}, L.~M., {Ade}, P. A.~R., {Angil{\`e}}, F.~E., {et~al.} 2019, \apj,
  878, 110, \dodoi{10.3847/1538-4357/ab1eb0}

\bibitem[{{Girart} {et~al.}(2009){Girart}, {Beltr{\'a}n}, {Zhang}, {Rao}, \&
  {Estalella}}]{Girart09}
{Girart}, J.~M., {Beltr{\'a}n}, M.~T., {Zhang}, Q., {Rao}, R., \& {Estalella},
  R. 2009, Science, 324, 1408, \dodoi{10.1126/science.1171807}

\bibitem[{{Goddi} {et~al.}(2020){Goddi}, {Ginsburg}, {Maud}, {Zhang}, \&
  {Zapata}}]{Goddi20}
{Goddi}, C., {Ginsburg}, A., {Maud}, L.~T., {Zhang}, Q., \& {Zapata}, L.~A.
  2020, \apj, 905, 25, \dodoi{10.3847/1538-4357/abc88e}

\bibitem[{{Hoq} {et~al.}(2013){Hoq}, {Jackson}, {Foster}, {Sanhueza},
  {Guzm{\'a}n}, {Whitaker}, {Claysmith}, {Rathborne}, {Vasyunina}, \&
  {Vasyunin}}]{Hoq13}
{Hoq}, S., {Jackson}, J.~M., {Foster}, J.~B., {et~al.} 2013, \apj, 777, 157,
  \dodoi{10.1088/0004-637X/777/2/157}

\bibitem[{{Hull} \& {Zhang}(2019)}]{Hull19}
{Hull}, C. L.~H., \& {Zhang}, Q. 2019, Frontiers in Astronomy and Space
  Sciences, 6, 3, \dodoi{10.3389/fspas.2019.00003}

\bibitem[{{Hull} {et~al.}(2017){Hull}, {Mocz}, {Burkhart}, {Goodman}, {Girart},
  {Cort{\'e}s}, {Hernquist}, {Springel}, {Li}, \& {Lai}}]{Hull17}
{Hull}, C. L.~H., {Mocz}, P., {Burkhart}, B., {et~al.} 2017, \apjl, 842, L9,
  \dodoi{10.3847/2041-8213/aa71b7}

\bibitem[{{Hull} {et~al.}(2020){Hull}, {Cortes}, {Gouellec}, {Girart}, {Nagai},
  {Nakanishi}, {Kameno}, {Fomalont}, {Brogan}, {Moellenbrock}, {Paladino}, \&
  {Villard}}]{Hull20}
{Hull}, C. L.~H., {Cortes}, P.~C., {Gouellec}, V. J.~M.~L., {et~al.} 2020,
  \pasp, 132, 094501, \dodoi{10.1088/1538-3873/ab99cd}

\bibitem[{{Izquierdo} {et~al.}(2018){Izquierdo}, {Galv{\'a}n-Madrid}, {Maud},
  {Hoare}, {Johnston}, {Keto}, {Zhang}, \& {de Wit}}]{Izquierdo18}
{Izquierdo}, A.~F., {Galv{\'a}n-Madrid}, R., {Maud}, L.~T., {et~al.} 2018,
  \mnras, 478, 2505, \dodoi{10.1093/mnras/sty1096}

\bibitem[{{Juvela} {et~al.}(2018){Juvela}, {Guillet}, {Liu}, {Ristorcelli},
  {Pelkonen}, {Alina}, {Bronfman}, {Eden}, {Kim}, {Koch}, {Kwon}, {Lee},
  {Malinen}, {Micelotta}, {Montillaud}, {Rawlings}, {Sanhueza}, {Soam},
  {Traficante}, {Ysard}, \& {Zhang}}]{Juvela18}
{Juvela}, M., {Guillet}, V., {Liu}, T., {et~al.} 2018, \aap, 620, A26,
  \dodoi{10.1051/0004-6361/201833245}

\bibitem[{{Koch} {et~al.}(2018){Koch}, {Tang}, {Ho}, {Yen}, {Su}, \&
  {Takakuwa}}]{Koch18}
{Koch}, P.~M., {Tang}, Y.-W., {Ho}, P. T.~P., {et~al.} 2018, \apj, 855, 39,
  \dodoi{10.3847/1538-4357/aaa4c1}

\bibitem[{{Koch} {et~al.}(2014){Koch}, {Tang}, {Ho}, {Zhang}, {Girart}, {Chen},
  {Frau}, {Li}, {Li}, {Liu}, {Padovani}, {Qiu}, {Yen}, {Chen}, {Ching}, {Lai},
  \& {Rao}}]{Koch14}
---. 2014, \apj, 797, 99, \dodoi{10.1088/0004-637X/797/2/99}

\bibitem[{{Lampton} {et~al.}(1976){Lampton}, {Margon}, \& {Bowyer}}]{Lampton76}
{Lampton}, M., {Margon}, B., \& {Bowyer}, S. 1976, \apj, 208, 177,
  \dodoi{10.1086/154592}

\bibitem[{{Li} \& {Shu}(1996)}]{Li96}
{Li}, Z.-Y., \& {Shu}, F.~H. 1996, \apj, 472, 211, \dodoi{10.1086/178056}

\bibitem[{{Liu} {et~al.}(2015){Liu}, {Galv{\'a}n-Madrid}, {Jim{\'e}nez-Serra},
  {Rom{\'a}n-Z{\'u}{\~n}iga}, {Zhang}, {Li}, \& {Chen}}]{Liu15}
{Liu}, H.~B., {Galv{\'a}n-Madrid}, R., {Jim{\'e}nez-Serra}, I., {et~al.} 2015,
  \apj, 804, 37, \dodoi{10.1088/0004-637X/804/1/37}

\bibitem[{{Liu} {et~al.}(2018){Liu}, {Li}, {Juvela}, {Kim}, {Evans}, {Di
  Francesco}, {Liu}, {Yuan}, {Tatematsu}, {Zhang}, {Ward-Thompson}, {Fuller},
  {Goldsmith}, {Koch}, {Sanhueza}, {Ristorcelli}, {Kang}, {Chen}, {Hirano},
  {Wu}, {Sokolov}, {Lee}, {White}, {Wang}, {Eden}, {Li}, {Thompson}, {Pattle},
  {Soam}, {Nasedkin}, {Kim}, {Kim}, {Lai}, {Park}, {Qiu}, {Zhang}, {Alina},
  {Eswaraiah}, {Falgarone}, {Fich}, {Greaves}, {Gu}, {Kwon}, {Li}, {Malinen},
  {Montier}, {Parsons}, {Qin}, {Rawlings}, {Ren}, {Tang}, {Tang}, {Toth},
  {Wang}, {Wouterloot}, {Yi}, \& {Zhang}}]{Liu18}
{Liu}, T., {Li}, P.~S., {Juvela}, M., {et~al.} 2018, \apj, 859, 151,
  \dodoi{10.3847/1538-4357/aac025}

\bibitem[{{Lu} {et~al.}(2014){Lu}, {Zhang}, {Liu}, {Wang}, \& {Gu}}]{Lu14}
{Lu}, X., {Zhang}, Q., {Liu}, H.~B., {Wang}, J., \& {Gu}, Q. 2014, \apj, 790,
  84, \dodoi{10.1088/0004-637X/790/2/84}

\bibitem[{{Lu} {et~al.}(2018){Lu}, {Zhang}, {Liu}, {Sanhueza}, {Tatematsu},
  {Feng}, {Smith}, {Myers}, {Sridharan}, \& {Gu}}]{Lu18}
{Lu}, X., {Zhang}, Q., {Liu}, H.~B., {et~al.} 2018, \apj, 855, 9,
  \dodoi{10.3847/1538-4357/aaad11}

\bibitem[{{Machida} {et~al.}(2005){Machida}, {Matsumoto}, {Tomisaka}, \&
  {Hanawa}}]{Machida05}
{Machida}, M.~N., {Matsumoto}, T., {Tomisaka}, K., \& {Hanawa}, T. 2005,
  \mnras, 362, 369, \dodoi{10.1111/j.1365-2966.2005.09297.x}

\bibitem[{{Maud} {et~al.}(2017){Maud}, {Hoare}, {Galv{\'a}n-Madrid}, {Zhang},
  {de Wit}, {Keto}, {Johnston}, \& {Pineda}}]{Maud17}
{Maud}, L.~T., {Hoare}, M.~G., {Galv{\'a}n-Madrid}, R., {et~al.} 2017, \mnras,
  467, L120, \dodoi{10.1093/mnrasl/slx010}

\bibitem[{{McKee} \& {Zweibel}(1992)}]{McKee92}
{McKee}, C.~F., \& {Zweibel}, E.~G. 1992, \apj, 399, 551,
  \dodoi{10.1086/171946}

\bibitem[{{McMullin} {et~al.}(2007){McMullin}, {Waters}, {Schiebel}, {Young},
  \& {Golap}}]{McMullin07}
{McMullin}, J.~P., {Waters}, B., {Schiebel}, D., {Young}, W., \& {Golap}, K.
  2007, in Astronomical Society of the Pacific Conference Series, Vol. 376,
  Astronomical Data Analysis Software and Systems XVI, ed. R.~A. {Shaw},
  F.~{Hill}, \& D.~J. {Bell}, 127

\bibitem[{{Olguin} {et~al.}(2021){Olguin}, {Sanhueza}, {Guzm{\'a}n}, {Lu},
  {Saigo}, {Zhang}, {Silva}, {Chen}, {Li}, {Ohashi}, {Nakamura}, {Sakai}, \&
  {Wu}}]{Olguin21}
{Olguin}, F.~A., {Sanhueza}, P., {Guzm{\'a}n}, A.~E., {et~al.} 2021, \apj, 909,
  199, \dodoi{10.3847/1538-4357/abde3f}

\bibitem[{{Ossenkopf} \& {Henning}(1994)}]{OH94}
{Ossenkopf}, V., \& {Henning}, T. 1994, \aap, 291, 943

\bibitem[{{Ostriker} {et~al.}(2001){Ostriker}, {Stone}, \&
  {Gammie}}]{Ostriker01}
{Ostriker}, E.~C., {Stone}, J.~M., \& {Gammie}, C.~F. 2001, \apj, 546, 980,
  \dodoi{10.1086/318290}

\bibitem[{{Padovani} \& {Galli}(2011)}]{Padovani11}
{Padovani}, M., \& {Galli}, D. 2011, \aap, 530, A109,
  \dodoi{10.1051/0004-6361/201116853}

\bibitem[{{Padovani} {et~al.}(2013){Padovani}, {Hennebelle}, \&
  {Galli}}]{Padovani13}
{Padovani}, M., {Hennebelle}, P., \& {Galli}, D. 2013, \aap, 560, A114,
  \dodoi{10.1051/0004-6361/201322407}

\bibitem[{{Padovani} {et~al.}(2012){Padovani}, {Brinch}, {Girart},
  {J{\o}rgensen}, {Frau}, {Hennebelle}, {Kuiper}, {Vlemmings}, {Bertoldi},
  {Hogerheijde}, {Juhasz}, \& {Schaaf}}]{Padovani12}
{Padovani}, M., {Brinch}, C., {Girart}, J.~M., {et~al.} 2012, \aap, 543, A16,
  \dodoi{10.1051/0004-6361/201219028}

\bibitem[{{Palmeirim} {et~al.}(2013){Palmeirim}, {Andr{\'e}}, {Kirk},
  {Ward-Thompson}, {Arzoumanian}, {K{\"o}nyves}, {Didelon}, {Schneider},
  {Benedettini}, {Bontemps}, {Di Francesco}, {Elia}, {Griffin}, {Hennemann},
  {Hill}, {Martin}, {Men'shchikov}, {Molinari}, {Motte}, {Nguyen Luong},
  {Nutter}, {Peretto}, {Pezzuto}, {Roy}, {Rygl}, {Spinoglio}, \&
  {White}}]{Palmeirim13}
{Palmeirim}, P., {Andr{\'e}}, P., {Kirk}, J., {et~al.} 2013, \aap, 550, A38,
  \dodoi{10.1051/0004-6361/201220500}

\bibitem[{{Planck Collaboration} {et~al.}(2016){Planck Collaboration}, {Ade},
  {Aghanim}, {Alves}, {Arnaud}, {Arzoumanian}, {Ashdown}, {Aumont},
  {Baccigalupi}, {Banday}, {Barreiro}, {Bartolo}, {Battaner}, {Benabed},
  {Beno{\^\i}t}, {Benoit-L{\'e}vy}, {Bernard}, {Bersanelli}, {Bielewicz},
  {Bock}, {Bonavera}, {Bond}, {Borrill}, {Bouchet}, {Boulanger}, {Bracco},
  {Burigana}, {Calabrese}, {Cardoso}, {Catalano}, {Chiang}, {Christensen},
  {Colombo}, {Combet}, {Couchot}, {Crill}, {Curto}, {Cuttaia}, {Danese},
  {Davies}, {Davis}, {de Bernardis}, {de Rosa}, {de Zotti}, {Delabrouille},
  {Dickinson}, {Diego}, {Dole}, {Donzelli}, {Dor{\'e}}, {Douspis}, {Ducout},
  {Dupac}, {Efstathiou}, {Elsner}, {En{\ss}lin}, {Eriksen},
  {Falceta-Gon{\c{c}}alves}, {Falgarone}, {Ferri{\`e}re}, {Finelli}, {Forni},
  {Frailis}, {Fraisse}, {Franceschi}, {Frejsel}, {Galeotta}, {Galli}, {Ganga},
  {Ghosh}, {Giard}, {Gjerl{\o}w}, {Gonz{\'a}lez-Nuevo}, {G{\'o}rski},
  {Gregorio}, {Gruppuso}, {Gudmundsson}, {Guillet}, {Harrison}, {Helou},
  {Hennebelle}, {Henrot-Versill{\'e}}, {Hern{\'a}ndez-Monteagudo}, {Herranz},
  {Hildebrandt}, {Hivon}, {Holmes}, {Hornstrup}, {Huffenberger}, {Hurier},
  {Jaffe}, {Jaffe}, {Jones}, {Juvela}, {Keih{\"a}nen}, {Keskitalo}, {Kisner},
  {Knoche}, {Kunz}, {Kurki-Suonio}, {Lagache}, {Lamarre}, {Lasenby},
  {Lattanzi}, {Lawrence}, {Leonardi}, {Levrier}, {Liguori}, {Lilje},
  {Linden-V{\o}rnle}, {L{\'o}pez-Caniego}, {Lubin}, {Mac{\'\i}as-P{\'e}rez},
  {Maino}, {Mandolesi}, {Mangilli}, {Maris}, {Martin},
  {Mart{\'\i}nez-Gonz{\'a}lez}, {Masi}, {Matarrese}, {Melchiorri}, {Mendes},
  {Mennella}, {Migliaccio}, {Miville-Desch{\^e}nes}, {Moneti}, {Montier},
  {Morgante}, {Mortlock}, {Munshi}, {Murphy}, {Naselsky}, {Nati},
  {Netterfield}, {Noviello}, {Novikov}, {Novikov}, {Oppermann}, {Oxborrow},
  {Pagano}, {Pajot}, {Paladini}, {Paoletti}, {Pasian}, {Perotto}, {Pettorino},
  {Piacentini}, {Piat}, {Pierpaoli}, {Pietrobon}, {Plaszczynski},
  {Pointecouteau}, {Polenta}, {Ponthieu}, {Pratt}, {Prunet}, {Puget}, {Rachen},
  {Reinecke}, {Remazeilles}, {Renault}, {Renzi}, {Ristorcelli}, {Rocha},
  {Rossetti}, {Roudier}, {Rubi{\~n}o-Mart{\'\i}n}, {Rusholme}, {Sandri},
  {Santos}, {Savelainen}, {Savini}, {Scott}, {Soler}, {Stolyarov}, {Sudiwala},
  {Sutton}, {Suur-Uski}, {Sygnet}, {Tauber}, {Terenzi}, {Toffolatti}, {Tomasi},
  {Tristram}, {Tucci}, {Umana}, {Valenziano}, {Valiviita}, {Van Tent},
  {Vielva}, {Villa}, {Wade}, {Wandelt}, {Wehus}, {Ysard}, {Yvon}, \&
  {Zonca}}]{Planck16}
{Planck Collaboration}, {Ade}, P.~A.~R., {Aghanim}, N., {et~al.} 2016, \aap,
  586, A138, \dodoi{10.1051/0004-6361/201525896}

\bibitem[{{Sanhueza} {et~al.}(2012){Sanhueza}, {Jackson}, {Foster}, {Garay},
  {Silva}, \& {Finn}}]{Sanhueza12}
{Sanhueza}, P., {Jackson}, J.~M., {Foster}, J.~B., {et~al.} 2012, \apj, 756,
  60, \dodoi{10.1088/0004-637X/756/1/60}

\bibitem[{{Sanhueza} {et~al.}(2017){Sanhueza}, {Jackson}, {Zhang},
  {Guzm{\'a}n}, {Lu}, {Stephens}, {Wang}, \& {Tatematsu}}]{Sanhueza17}
{Sanhueza}, P., {Jackson}, J.~M., {Zhang}, Q., {et~al.} 2017, \apj, 841, 97,
  \dodoi{10.3847/1538-4357/aa6ff8}

\bibitem[{{Soam} {et~al.}(2019){Soam}, {Liu}, {Andersson}, {Lee}, {Liu},
  {Juvela}, {Li}, {Goldsmith}, {Zhang}, {Koch}, {Kim}, {Qiu}, {Evans},
  {Johnstone}, {Thompson}, {Ward-Thompson}, {Di Francesco}, {Tang},
  {Montillaud}, {Kim}, {Mairs}, {Sanhueza}, {Kim}, {Berry}, {Gordon},
  {Tatematsu}, {Liu}, {Pattle}, {Eden}, {McGehee}, {Wang}, {Ristorcelli},
  {Graves}, {Alina}, {Lacaille}, {Montier}, {Park}, {Kwon}, {Chung},
  {Pelkonen}, {Micelotta}, {Saajasto}, \& {Fuller}}]{Soam19}
{Soam}, A., {Liu}, T., {Andersson}, B.~G., {et~al.} 2019, \apj, 883, 95,
  \dodoi{10.3847/1538-4357/ab39dd}

\bibitem[{{Soler} {et~al.}(2017){Soler}, {Ade}, {Angil{\`e}}, {Ashton},
  {Benton}, {Devlin}, {Dober}, {Fissel}, {Fukui}, {Galitzki}, {Gandilo},
  {Hennebelle}, {Klein}, {Li}, {Korotkov}, {Martin}, {Matthews}, {Moncelsi},
  {Netterfield}, {Novak}, {Pascale}, {Poidevin}, {Santos}, {Savini}, {Scott},
  {Shariff}, {Thomas}, {Tucker}, {Tucker}, \& {Ward-Thompson}}]{Soler17}
{Soler}, J.~D., {Ade}, P.~A.~R., {Angil{\`e}}, F.~E., {et~al.} 2017, \aap, 603,
  A64, \dodoi{10.1051/0004-6361/201730608}

\bibitem[{{Sridharan} {et~al.}(2002){Sridharan}, {Beuther}, {Schilke},
  {Menten}, \& {Wyrowski}}]{Sridharan02}
{Sridharan}, T.~K., {Beuther}, H., {Schilke}, P., {Menten}, K.~M., \&
  {Wyrowski}, F. 2002, \apj, 566, 931, \dodoi{10.1086/338332}

\bibitem[{{Tobin} {et~al.}(2012){Tobin}, {Hartmann}, {Bergin}, {Chiang},
  {Looney}, {Chandler}, {Maret}, \& {Heitsch}}]{Tobin12}
{Tobin}, J.~J., {Hartmann}, L., {Bergin}, E., {et~al.} 2012, \apj, 748, 16,
  \dodoi{10.1088/0004-637X/748/1/16}

\bibitem[{{Vaillancourt}(2006)}]{Vaillancourt06}
{Vaillancourt}, J.~E. 2006, \pasp, 118, 1340, \dodoi{10.1086/507472}

\bibitem[{{Vlemmings}(2008)}]{Vlemmings08}
{Vlemmings}, W.~H.~T. 2008, \aap, 484, 773, \dodoi{10.1051/0004-6361:200809447}

\bibitem[{{Xu} {et~al.}(2011){Xu}, {Moscadelli}, {Reid}, {Menten}, {Zhang},
  {Zheng}, \& {Brunthaler}}]{Xu11}
{Xu}, Y., {Moscadelli}, L., {Reid}, M.~J., {et~al.} 2011, \apj, 733, 25,
  \dodoi{10.1088/0004-637X/733/1/25}

\bibitem[{{Zapata} {et~al.}(2006){Zapata}, {Rodr{\'\i}guez}, {Ho}, {Beuther},
  \& {Zhang}}]{Zapata06}
{Zapata}, L.~A., {Rodr{\'\i}guez}, L.~F., {Ho}, P. T.~P., {Beuther}, H., \&
  {Zhang}, Q. 2006, \aj, 131, 939, \dodoi{10.1086/499156}

\bibitem[{{Zhang} {et~al.}(2014){Zhang}, {Qiu}, {Girart}, {Liu}, {Tang},
  {Koch}, {Li}, {Keto}, {Ho}, {Rao}, {Lai}, {Ching}, {Frau}, {Chen}, {Li},
  {Padovani}, {Bontemps}, {Csengeri}, \& {Ju{\'a}rez}}]{Zhang14}
{Zhang}, Q., {Qiu}, K., {Girart}, J.~M., {et~al.} 2014, \apj, 792, 116,
  \dodoi{10.1088/0004-637X/792/2/116}

\end{thebibliography}
\bibliographystyle{aasjournal}



\end{document}